\documentclass[aps,twocolumn,superscriptaddress]{revtex4-1}
\usepackage[colorlinks=true,bookmarks=true,citecolor=magenta,urlcolor=magenta,linkcolor=magenta,breaklinks]{hyperref} 
\usepackage{breakurl}
\usepackage{sidecap}
\usepackage{amssymb}
\usepackage{hhline}
\usepackage{urwchancal}
\usepackage{xcolor}
\sidecaptionvpos{figure}{t}
\usepackage{amsmath}
\usepackage{graphicx}
\usepackage{esint}
\usepackage{epstopdf}
\usepackage{rotating}
\graphicspath{{pict/}{./}}
\usepackage{bm}%
\usepackage{txfonts}
\usepackage[classicReIm]{kpfonts}

\newcounter{Fig}

\begin{document}


\title{Electromagnetic duality protected scattering properties of nonmagnetic particles }

\author{Qingdong Yang}
\email{Authors contributed equally to this work.}
\affiliation{School of Optical and Electronic Information, Huazhong University of Science and Technology, Wuhan, Hubei 430074, P. R. China}
\author{Weijin Chen}
\email{Authors contributed equally to this work.}
\affiliation{School of Optical and Electronic Information, Huazhong University of Science and Technology, Wuhan, Hubei 430074, P. R. China}
\author{Yuntian Chen}
\email{yuntian@hust.edu.cn}
\affiliation{School of Optical and Electronic Information, Huazhong University of Science and Technology, Wuhan, Hubei 430074, P. R. China}
\affiliation{Wuhan National Laboratory for Optoelectronics, Huazhong University of Science and Technology, Wuhan, Hubei 430074, P. R. China}
\author{Wei Liu}
\email{wei.liu.pku@gmail.com}
\affiliation{College for Advanced Interdisciplinary Studies, National University of Defense
Technology, Changsha, Hunan 410073, P. R. China}

\begin{abstract}
Optical properties of nonmagnetic structures that support artificial optically-induced magnetic responses have recently attracted surging interest. Here we conduct symmetry-dictated investigations into scattering properties of nonmagnetic particles from perspectives of electromagnetic duality with discrete geometric rotations. For arbitrary scattering configurations, we reveal that far-field scattering patterns are invariant under duality transformations, which in particular means that scattering patterns of self-dual clusters with random particle distributions are polarization independent. Based on this revelation, it is further discovered that scattering bodies of combined duality-(n-fold) rotation symmetry, for any polarizations of incident waves, exhibit also n-fold rotationally symmetric scattering patterns with zero backward components, satisfying the first Kerker condition automatically. We employ both coupled dipole theory and full numerical simulations to demonstrate those scattering properties,  solely based upon nonmagnetic core-shell particles that support optically-induced dipolar resonances. Those substantiated scattering properties are fully induced by fundamental symmetry principles, and thus can survive any non-symmetry-breaking perturbations, which may find applications in a wide range of optical devices that require intrinsically robust functionalities.
\end{abstract}

\maketitle

\section{Introduction}
\label{section1}

The fundamental concept of optically-induced magnetism in nonmagnetic structures~\cite{Pendry1999_ITMT} has been playing a central role in the field of metamaterials and metasurfaces~\cite{Cai2010_book,CHEN_Rep.Prog.Phys._review_2016}, constantly fertilising many other branches of optics and photonics~\cite{jahani_alldielectric_2016,KUZNETSOV_Science_optically_2016,OZAWA_2018_ArXiv180204173}. Basically, the introduction of optically-induced magnetic responses renders an extra degree of freedom for manipulations of light-matter interactions, through not only the electric but also the magnetic components of electromagnetic waves. This has in particular triggered enormous renewed interest in the seminal problem of electromagnetic scatterings, mainly through Kerker scattering and its generalized forms~\cite{Kerker1983_JOSA,LIU_2018_Opt.Express_Generalized}. The artificial optical magnetism enables thorough exploitations of interferences between simultaneous electric and  magnetic resonances of different orders, which makes possible many applications that are inaccessible to sole electric responses~\cite{CHEN_Rep.Prog.Phys._review_2016,jahani_alldielectric_2016,KUZNETSOV_Science_optically_2016,OZAWA_2018_ArXiv180204173}.

It is well known that, until the recent reinvigoration rendered by optically-induced magnetism~\cite{LIU_2018_Opt.Express_Generalized}, Kerker's original proposal based on hypothetical magnetic particles to eliminate backward scatterings had not attracted much attention~\cite{Kerker1983_JOSA}. This is mainly due to the fact that too few materials are intrinsically magnetic, especially at high frequency spectral regimes. Exactly the same obstacle stays in the way for widely employing the fundamental electromagnetic duality principle for photonic researches and applications, which in a similar fashion requires simultaneous electric and magnetic responses~\cite{jackson1962classical,FERNANDEZ-CORBATON_2013_Phys.Rev.Lett._Electromagnetica}. It is therefore expected that the combination of optically-induced magnetism and electromagnetic duality would render fresh opportunities for manipulating optical properties of scattering particles, setting free electromagnetic duality related fundamental studies and applications from the rather rare intrinsically magnetic materials.

In this work, we study the plane wave scatterings by nonmagnetic particle clusters that support simultaneous electric and optically-induced magnetic responses. This study is solely dictated by symmetry principles involving duality transformations and discrete geometric rotations. We reveal that far-field scattering patterns for arbitrary particle clusters are invariant under duality transformations: as a special case of this, the scattering patterns of random self-dual particle clusters are polarization independent.  Based on this discovery, it is further revealed that the combined duality-(n-fold) rotation symmetry would guarantee: (i) the scattering patterns also exhibit n-fold rotation symmetry; (ii) the backward scatterings are intrinsically eliminated, with the first Kerker condition automatically satisfied. We have verified those scattering properties through both analytical coupled dipole theory and full numerical simulations, employing realistic nonmagnetic core-shell particles. The scattering properties we have revealed are symmetry protected, which are independent of specific geometric parameters and also immune to perturbations and inter-particle couplings that preserve the required symmetry. Our explorations may find applications in the rapidly expanding fields of topological and/or non-hermitian photonics, bring new opportunities for a wide range of optical systems that require robust functionalities.

\section{Duality transformations for isotropic structures without magneto-electric couplings}
\label{section2}
 Electromagnetic duality transformations are defined as (in Gaussian units; \textbf{E}, \textbf{H}, \textbf{D} and \textbf{B} are widely adopted notations that correspond to usual electromagnetic quantities)~\cite{jackson1962classical,FERNANDEZ-CORBATON_2013_Phys.Rev.Lett._Electromagnetica}:

\begin{equation}
\label{duality_transformation}
\begin{aligned}
&\left(\begin{array}{c}
\mathrm{\mathbf{E}} \\
\mathbf{H}
\end{array}\right) \rightarrow\left(\begin{array}{l}
\mathrm{\mathbf{E}}^{\prime} \\
\mathbf{H}^{\prime}
\end{array}\right)=T(\beta)\left(\begin{array}{l}
\mathrm{\mathbf{E}} \\
\mathbf{H}
\end{array}\right),\\
&\left(\begin{array}{l}
\mathrm{\mathbf{D}} \\
\mathbf{B}
\end{array}\right) \rightarrow\left(\begin{array}{l}
\mathrm{\mathbf{D}}^{\prime} \\
\mathbf{B}^{\prime}
\end{array}\right)=T(\beta)\left(\begin{array}{l}
\mathrm{\mathbf{D}} \\
\mathbf{B}
\end{array}\right).
\end{aligned}
\end{equation}
Here $T(\beta)$ is the duality transformation matrix:
\begin{equation}
\label{duality_rotation_matrix}
T(\beta)=\left[\begin{array}{ll}
\cos \beta, & -\sin \beta\\
\sin \beta, & ~~\cos \beta\end{array}\right],\end{equation}
and $\beta$ is a real transformation angle. It is crucial  to mention that though $T(\beta)$ looks to be identical to the two-dimensional (2D) rigid rotation matrix $R(\beta)$:
\begin{equation}
\label{duality_rotation_matrix}
R(\beta)=\left[\begin{array}{ll}
\cos \beta, & -\sin \beta\\
\sin \beta, & ~~\cos \beta\end{array}\right],\end{equation}
the two matrices correspond to fundamentally different operations. Though $T(\beta)$ is usually termed as a duality rotation matrix, it does not usually represent a rigourously geometric rotation since for electromagnetic waves, \textbf{E} (\textbf{D}) are not generally perpendicular to \textbf{H} (\textbf{B}). Even if this is the case (\textit{e.g.} plane waves) and  $T(\beta)$ can be effectively viewed as a geometric rotation, such a rotation is still different from $R(\beta)$: $T(\beta)$ corresponds to a local rotation,  the rotation axis of which (along $\mathbf{E\times H}$ or $\mathbf{D\times B}$) is position-dependent and decided by the spatial field distributions; while $R(\beta)$ corresponds to a global rotation, the rotation axis of which is spatially fixed. With the absence of free electric charge and currents, electromagnetic duality means the invariance of Maxwell equations under the above transformations in Eq.~(\ref{duality_transformation}), ensuring that if one of the two sets of fields (\textbf{E}, \textbf{H}, \textbf{D} \textbf{B}) and (\textbf{E}$^\prime$, \textbf{H}$^\prime$, \textbf{D}$^\prime$, \textbf{B}$^\prime$) is the solution of the Maxwell equation, the other must be a solution too.

When we confine our studies to isotropic materials of permittivity $\epsilon$ and permeability $\mu$ without magneto-electric couplings, the field components are connected through~\cite{jackson1962classical}:
 \begin{equation}
 \label{D-B}
 \left[\begin{array}{l}
\mathbf{D} \\
\mathbf{B}
\end{array}\right]=\left[\begin{array}{ll}
\epsilon, & 0 \\
0, & \mu
\end{array}\right]\left[\begin{array}{l}
\mathbf{E} \\
\mathbf{H}
\end{array}\right].\end{equation}
Together with Eq.~(\ref{duality_transformation}) it leads to:
\begin{equation}
\label{bi-anisotropy-pre}
\left[\begin{array}{l}
\mathbf{D}^{\prime} \\
\mathbf{B}^{\prime}
\end{array}\right]=T(\beta)\left[\begin{array}{ll}
\epsilon, & 0 \\
0, & \mu
\end{array}\right] T(-\beta)\left[\begin{array}{l}
\mathbf{E}^{\prime} \\
\mathbf{H}^{\prime}
\end{array}\right],\end{equation}
which can be more explicitly expressed as:
 \begin{equation}
 \label{bi-anisotropy}
 \left[\begin{array}{l}
\mathbf{D^\prime}\\
\mathbf{B^\prime}
\end{array}\right]=\left[\begin{array}{ll}
\epsilon+(\mu-\epsilon)\sin^2\beta,&\frac{1}{2}(\epsilon-\mu)\sin(2\beta) \\
\frac{1}{2}(\epsilon-\mu)\sin(2\beta),&\epsilon+(\mu-\epsilon)\cos^2\beta
\end{array}\right]\left[\begin{array}{l}
\mathbf{E^\prime}\\
\mathbf{H^\prime}
\end{array}\right].\end{equation}
Equation~(\ref{bi-anisotropy}) shows that general duality transformations requires non-reciprocal bi-isotropic materials within which electric and magnetic fields are coupled with each other~\cite{SIHVOLA_1994__Electromagnetic}. Such materials are intrinsically magnetic and nonreciprocal, and thus cannot be realized through optically-induced magnetism that does not break reciprocity. If there are no magneto-electric couplings, the off diagonal terms in the matrix of Eq.~(\ref{bi-anisotropy}) are zero: $(\epsilon-\mu)\sin(2\beta)=0$, which requires either $\epsilon=\mu$ or $\beta=\pi/2$ (other possible angles such as $\beta=3\pi/2$ are physically equivalent). For the former case of $\epsilon=\mu$, the materials are self-dual, for which duality transformations of arbitrary $\beta$ can be directly implemented, without changing the optical parameters of the structures; while the latter case  corresponds to a single duality transformation of $(\mathbf{E,D}) \rightarrow (\mathbf{H, B}$), $(\mathbf{H,B}) \rightarrow (\mathbf{-E,-D})$, for which a simultaneous parameter interchange $(\mu,\epsilon) \rightarrow (\epsilon, \mu)$ is required.

\section{Invariance of scattering patterns under duality transformations}
\label{section-3}

Now we proceed to the seminal problem of plane wave scatterings by obstacles in free space (which we assume to be vacuum in this study), and study general scattering properties from the perspective of duality transformations. As has been discussed in Section~$\rm{\ref{section2}}$, for electromagnetic waves with perpendicular electric and magnetic fields, duality transformations can be viewed as 2D geometric rotations (duality rotations) with the rotation axes parallel to the propagation direction (Poynting vector). This geometric interpretation can be employed for not only incident plane waves, but also  far-field scattered waves which are effectively plane waves propagating along different scattering directions. For incident plane waves, duality transformations correspond to global rigid rotations of the fields and thus also polarizations by an angle $\beta$ [refer to Eqs.~(\ref{duality_transformation})-(\ref{duality_rotation_matrix})]; while for far-field scattered waves, duality transformations correspond to local rotations of fields, the rotation axes of which are along the corresponding scattering directions. Since the duality rotation changes neither the field magnitudes nor their propagation directions (it changes only the field orientations), the far-field scattering patterns (in terms of either field intensity or energy flux) are invariant under duality transformations.

Throughout this work, though we discuss only the invariance of scattering patterns, it should be reminded that actually all the cross sections of extinction, scattering and absorption are also invariant under duality transformations considering the following arguments: (i) The scattering cross section is solely decided the integration of the angular scatterings, and thus the invariance of scattering patterns would directly result in the invariance of scattering cross sections; (ii) According to the optical theorem~\cite{Bohren1983_book}, the extinction cross section originates from the interference of the incident wave and the forward scattered waves. Since a duality transformation  corresponds to an identical geometric rotation of both waves, the interference is the same and thus the extinction cross section is also invariant; (iii) The invariance of both scattering and extinction cross sections would directly guarantee the invariance of absorption cross sections, since absorption is nothing but extinction minus scattering as told also by the optical theorem.

\subsection{Duality transformations for dipolar Mie particles}

To verify the scattering pattern invariance, we turn to nonmagnetic Mie particles that can support a pair of electric and optically-induced magnetic dipolar moments~\cite{Wheeler2006_PRB,Liu2012_ACSNANO} for detailed demonstrations. For spherical particles  with sole dipolar resonances, the effective electric and magnetic isotropic polarizabilities are~\cite{Wheeler2006_PRB,Liu2012_ACSNANO,Bohren1983_book}:
\begin{equation}
\label{polarizabilities}
{\alpha^{e}} = {{3i} \over {2{k^3}}}{a_1},~{\alpha^{m}} = {{3i} \over {2{k^3}}}{b_1},
\end{equation}
where $a_1$ and $b_1$ are Mie scattering coefficients, which correspond respectively to electric dipole (ED) and magnetic dipole (MD) moments and can be calculated analytically (for both single and multi-layered structures)~\cite{Bohren1983_book}; and $k$ is the angular wave number in the background media. Then the electric and magnetic dipolar moments (\textbf{p} and \textbf{m}) can be expressed as:
\begin{equation}
\label{dipolar_moments}
\left(\begin{array}{c}
\mathbf{p} \\
\mathbf{m}
\end{array}\right)=\left[\begin{array}{cc}
\alpha^{e}, & 0 \\
0, & \alpha^{m}
\end{array}\right]\left(\begin{array}{c}
\mathbf{E} \\
\mathbf{H}
\end{array}\right).
\end{equation}
At the same time, the duality transformation for electric and magnetic moments is~\cite{jackson1962classical}:
\begin{equation}
\label{moment-transformation}
\left(\begin{array}{c}
\mathbf{p} \\
\mathbf{m}
\end{array}\right) \rightarrow\left(\begin{array}{c}
\mathbf{p}^{\prime} \\
\mathbf{m}^{\prime}
\end{array}\right)=T(\beta)\left(\begin{array}{c}
\mathbf{p} \\
\mathbf{m}
\end{array}\right).\end{equation}
Comparing Eqs.~(\ref{dipolar_moments})-(\ref{moment-transformation}) with Eqs.~(\ref{duality_transformation})-(\ref{D-B}), we can draw similar conclusions as those presented in Section~$\rm{\ref{section2}}$: duality transformations for isotropic spherical particles without magneto-electric couplings require either $\alpha^{m}=\alpha^{e}$ or $\beta=\pi/2$. For the former scenario of $\alpha^{m}=\alpha^{e}$, the particles are self-dual (as will be discussed later that they are their own dual-partners), for which duality transformations of arbitrary $\beta$ can be directly implemented, with the particles maintained as they are (in terms of both optical parameters and their geometric locations); The latter scenario ($\beta=\pi/2$) corresponds to a single duality transformation of $(\mathbf{E,p}) \rightarrow (\mathbf{H,m}$), $(\mathbf{H,m}) \rightarrow (\mathbf{-E,-p})$, and $(\alpha^{m},\alpha^{e}) \rightarrow (\alpha^{e},\alpha^{m})$, or equivalently $(a_{1},b_{1}) \rightarrow (b_{1},a_{1})$.

Both sets of duality transformations for isotropic responses without magneto-electric couplings can be directly confirmed through the coupled dipole theory (CDT) involving both electric and magnetic moments~\cite{Mulholland1994_Langmuir,Merchiers2007_PRA}.  For a cluster of  spherical particles (total particle number $N$) that support only dipolar responses,  within the $i$-th particle (Mie coefficients $a_{i1}$ and $b_{i1}$; electric and magnetic polarizabilities ${\alpha^{e}_i} = {{3i} \over {2{k^3}}}{a_{i1}},~{\alpha^{m}_i} = {{3i} \over {2{k^3}}}{b_{i1}}$) at the position $\mathbf{r}_i$ an electric moment $\mathbf{p}_i$ and magnetic moment $\mathbf{m}_i$ are excited. For an incident plane wave ($\mathbf{E}_i^0$ and $\mathbf{H}_i^0$ are the electric and magnetic fields at $\mathbf{r}_i$, respectively),  the coupled dipole equations are~\cite{Liu2012_ACSNANO,Mulholland1994_Langmuir,Merchiers2007_PRA}:
\begin{equation}
\label{coupled-dipole}
\begin{aligned}
&\mathbf{p}_{i}=\alpha_{i}^{e} \mathbf{E}_{i}^{0}+\alpha_{i}^{e} \sum_{j \neq i}\left(\mathbf{E}_{i}^{\mathbf{p}_{j}}+\mathbf{E}_{i}^{\mathbf{m}_{j}}\right),\\
&\mathbf{m}_{i}=\alpha_{i}^{m} \mathbf{H}_{i}^{0}+\alpha_{i}^{m} \sum_{j \neq i}\left(\mathbf{H}_{i}^{\mathbf{p}_{j}}+\mathbf{H}_{i}^{\mathbf{m}_{j}}\right),
\end{aligned}
\end{equation}
where $\mathbf{E}_i^{{\mathbf{p}_j}}$ and $\mathbf{H}_i^{{\mathbf{p}_j}}$ are electric and magnetic fields generated by the electric dipole $\mathbf{p}_j$ at $\mathbf{r}_i$, respectively;  and $\mathbf{E}_i^{{\mathbf{m}_j}}$ and $\mathbf{H}_i^{{\mathbf{m}_j}}$ are the electric and magnetic fields of the magnetic dipole $\mathbf{m}_j$ at $\mathbf{r}_i$, respectively. Simple algebraic manipulations can confirm that Eq.~(\ref{coupled-dipole}) are invariant under the duality transformation in Eqs.~(\ref{duality_transformation}) and (\ref{moment-transformation}): $\beta$ is arbitrary for self-dual particles $a_{i1}=b_{i1}$; for non-self dual particles $a_{i1}\neq b_{i1}$, it is required that $\beta=\pi/2$ and the Mie coefficient interchange $(a_{i1},b_{i1}) \rightarrow (b_{i1},a_{i1})$.  By solving Eq.~({\ref{coupled-dipole}), both sets of dipolar moments for all the particles can be calculated. With those moments at hand, the far-field scatterings at all angles can be obtained through superposing all the fields radiated by those dipolar moments~\cite{Liu2012_ACSNANO,Mulholland1994_Langmuir,Merchiers2007_PRA}.

\begin{figure}[tp]
\centerline{\includegraphics[width=8.8cm]{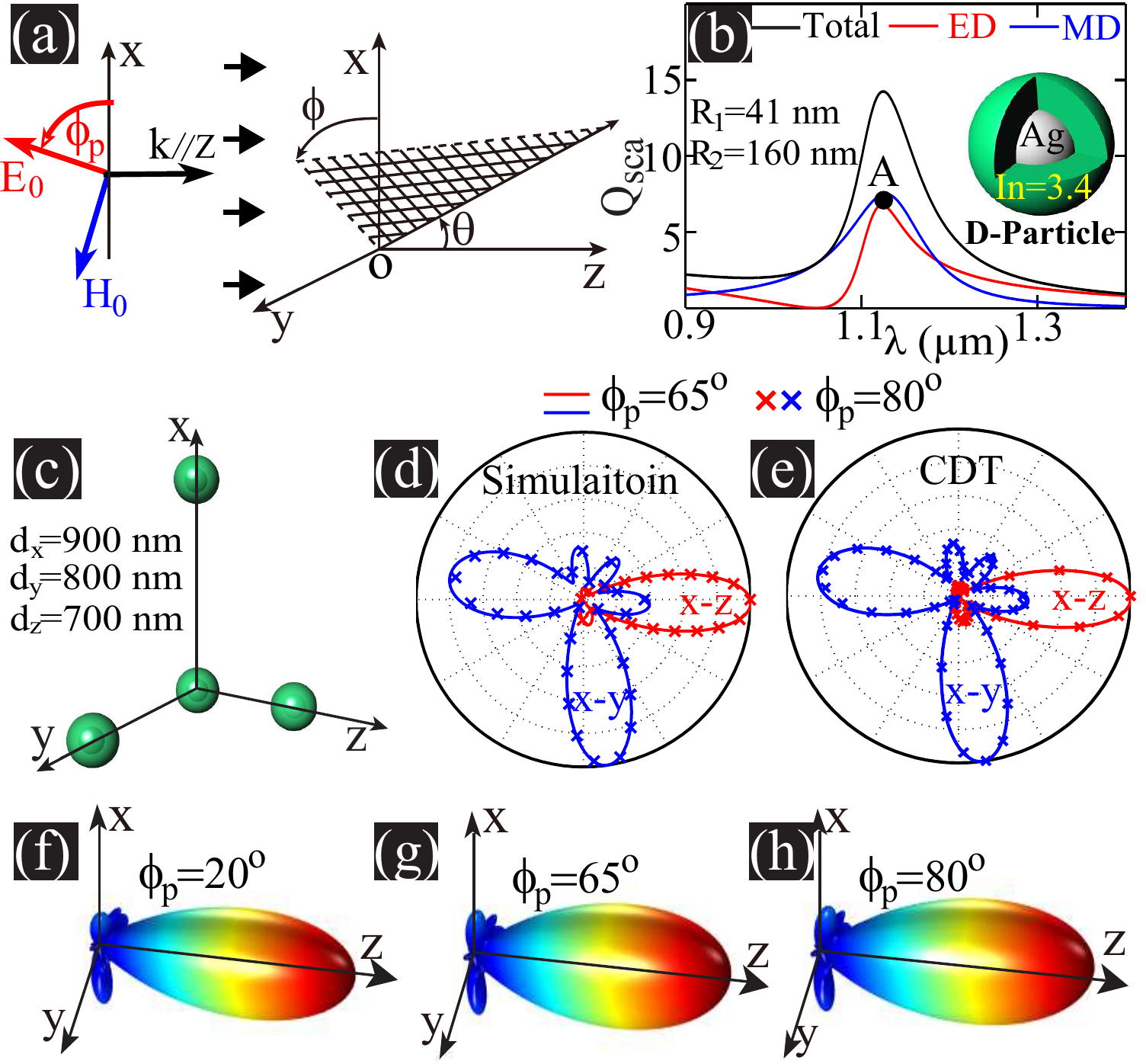}} \caption {\small (a) The Cartesian and polar coordinates with a shared origin. The plane wave is propagating along $\textbf{z}$, with the electric field polarized at the azimuthal angle $\phi_p$ (with respect to the $\textbf{x}$ axis). (b) Scattering efficiency spectra (both total scattering and partial dipolar contributions from ED and MD) for the Ag core-dielectric (In=$3.4$) shell spherical particle (see inset; $R_1=41$~nm and $R_2=160$~nm; termed as \textbf{D}-\textbf{Particle}) with the resonant position pinpointed at $\lambda_\mathbf{A}=1128$~nm. (c) A randomly chosen particle cluster consisting of four such \textbf{D}-\textbf{Particles}, with one at the origin and the other three at the three axes with displacements $d_x=900$~nm, $d_y=800$~nm, and $d_z=700$~nm. The 2D scattering patterns on both $\textbf{x-z}$ and $\textbf{x-y}$ planes for two polarizations ($\phi_p=~65^{\circ},~80^{\circ}$) at $\lambda_\mathbf{A}$ are shown in (d) and (e), where simulation and CDT results are included, respectively.  The corresponding 3D scattering patterns (with also results at an extra angle $\phi_p=~20^{\circ}$) through simulations are shown in (f)-(h).}\label{fig1}
\end{figure} 

\subsection{Polarization-independent scattering patterns of arbitrary clusters of self-dual particles}
Three points about plane wave scattering by self-dual dipolar particles ($a_{1}=b_{1}$) have been clarified in the discussions above: (i) For an incident plane wave, a duality transformation is equivalent to a geometric rotation by $\beta$ along its propagation direction, and thus both fields and the polarization are also rotated by $\beta$;  (ii) For arbitrary clusters of self-dual particles, duality transformations of arbitrary $\beta$ can be directly implemented with the particles maintained as they are; (iii) The scattering patterns are invariant under duality transformations. Those points direct lead to the following conclusion: For an arbitrary ensemble of self-dual particles, when the propagation direction of the incident wave is fixed, the scattering patterns are independent of the polarization directions.

To verify this polarization-independent scattering feature, we employ the self-dual metal (Ag)-dielectric (refractive index In=$3.4$) core-shell spherical particle that has been studied in previous studies~\cite{Liu2012_ACSNANO,Wheeler2006_PRB,Paniague2011_NJP}. The permittivity of silver is extracted from the experimental data in Ref.~\cite{Johnson1972_PRB}. Throughout this work, we place the scattering configuration in the coordinate system shown in Fig.~\ref{fig1}(a), where the plane wave is propagating along $\textbf{z}$ (wavevector $\textbf{k}\parallel \textbf{z}$)   and polarized on the $\mathbf{x-y}$ plane of an azimuthal polarization angle $\phi_p$. In Fig.~\ref{fig1}(b), we show the scattering efficiency spectra (in terms of total scattering and those contributed by ED ($a_1$) and MD ($b_1$)] for the particle of inner radius $R_1=41$~nm and outer radius $R_2=160$~nm [inset of Fig.~\ref{fig1}(b)].  As is shown, the particle is almost self-dual ( $a_1\approx b_1$) at the resonant wavelength $\lambda_\mathbf{A}=1128$~nm, and thus identified as \textbf{D}-\textbf{Particle} (here \textbf{D} means dual)  in this work.

For an arbitrary ensemble [one shown in Fig.~\ref{fig1}(c)] of \textbf{D}-\textbf{Particles} (refer to the caption of Fig.~\ref{fig1} for the randomly chosen position parameters), we show in Fig.~\ref{fig1}(d) (simulation results through COMSOL MULTIPHYSICS) and Fig.~\ref{fig1}(e) [CDT results through Eq.~(\ref{coupled-dipole})] the 2D scattering patterns on the $\textbf{x-z}$ and $\textbf{x-y}$ planes for two arbitrary polarizations ($\phi_p=65^{\circ}$ and $80^{\circ}$) at $\lambda_\mathbf{A}$. Throughout this work, for better visibility the 2D patterns on the two planes are normalized separately. Both sets of results agree with each other and can verify the independence of the scattering patterns on the polarizations. The minor deviations and discrepancies are induced either by:  $a_1$ and $b_1$ are not exactly equal to each other; there are some marginal contributions of quadrupoles that have not been taken into considerations by CDT.  We also show in Figs.~\ref{fig1}(f)-(h) the corresponding three-dimensional (3D) scattering patterns (simulation) for a further confirmation, with the results of an extra polarization angle $\phi_p=20^{\circ}$ also included. Here we show only a finite number of particles; for the infinite particle scenario, our arguments are certainly valid and thus polarization independence is also secured (see Ref.~\cite{Liu2012_PRB} for a detailed demonstration with Fano resonances).

\subsection{Invariance of scattering patterns under duality transformations for non-self-dual particles}

For non-self-dual dipolar particles ($a_{1}\neq b_{1}$) of isotropic responses without magneto-electric couplings,  only the duality transformation of $\beta=\pi/2$ is allowed, and such a transformation would convert dipolar electric (magnetic) moments to magnetic (electric) moments. To implement such a duality transformation, the original particles ($a_{1}=\kappa_1,~ b_{1}=\kappa_2$) should be replaced by their dual-partners ($a_{1}=\kappa_2,~ b_{1}=\kappa_1$), with the incident wave also rotated by $\pi/2$ along $\mathbf{k}$. This of course is also applicable to self-dual particles, as their dual-partners are themselves.

\begin{figure}[tp]
\centerline{\includegraphics[width=8.8cm]{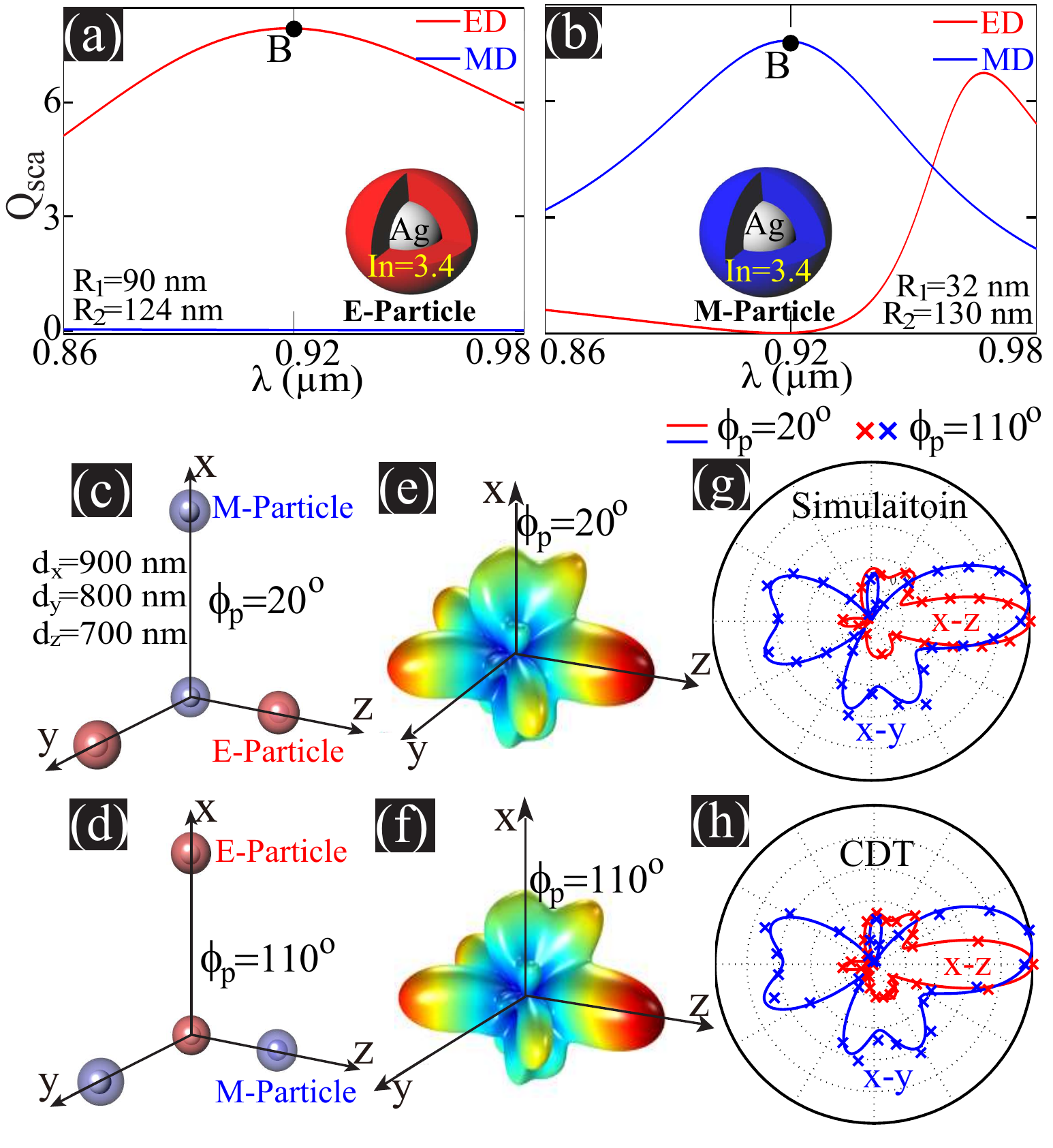}} \caption{\small (a) \& (b) Scattering efficiency spectra (dipolar contributions from ED and MD) for the Ag core-dielectric (In=$3.4$) shell spherical particles. At the common marked spectral point $\lambda_\mathbf{B}=920$~nm: \textbf{E}-\textbf{Particle} ($R_1=90$~nm and $R_2=124$~nm) supports pure electric moment as shown in (a); while \textbf{M}-\textbf{Particle} ($R_1=32$~nm and $R_2=130$~nm) supports pure magnetic moment as shown in (b). (c) \& (d) Two scattering configurations that are related to each other by a duality transformation of $\beta=\pi/2$ (interchanges between \textbf{E}-\textbf{Particles} and \textbf{M}-\textbf{Particle}). The two clusters shown are the same as that in Fig.~\ref{fig1}(c), except that the \textbf{D}-\textbf{Particles} are replaced by either \textbf{E}-\textbf{Particles} or \textbf{M}-\textbf{Particle}. The incident polarization angles are chosen as $\phi_p=20^{\circ}$ and $110^{\circ}$, respectively. The 2D scattering patterns (simulation and CDT results) on both $\textbf{x-z}$ and $\textbf{x-y}$ planes at $\lambda_\mathbf{B}$ are shown in (g) and (h), and the corresponding 3D scattering patterns are shown in (e) and (f).}
\label{fig2}
\end{figure}

The simplest dual-pairs would be \textbf{E-Particles} that support sole electric moments [$a_{1}(\mathbf{E})\neq 0,~ b_{1}(\mathbf{E})=0$] and \textbf{M-Particles} that support sole magnetic moments [$a_{1}(\mathbf{M})= 0,~ b_{1}(\mathbf{M})\neq 0$]~\cite{liu_toroidal_2015,FENG_2017_Phys.Rev.Lett._Ideal} when $a_{1}(\mathbf{E})=b_{1}(\mathbf{M})$.  We show scattering efficiencies of such a dual-pair: an \textbf{E-Particle} in Fig.~\ref{fig2}(a) ($R_1=90$~nm and $R_2=124$~nm) and a \textbf{M-Particle} in Fig.~\ref{fig2}(b) ($R_1=32$~nm and $R_2=130$~nm). Both particles are made of Ag-core and dielectric-shell (In=$3.4$) and support a pure (electric or magnetic) moment at the common wavelength $\lambda_\mathbf{B}=920$~nm. A randomly chosen cluster of such particles are shown in Fig.~\ref{fig2}(c), and its dual-pair cluster is shown in Fig.~\ref{fig2}(d), which is obtained through substituting all \textbf{E (M)-Particles} by dual-partner \textbf{M (E)-Particles}. To make sure the scattering configurations in Figs.~\ref{fig2}(c)-(d) are connected to each other by a duality transformation of $\beta=\pi/2$, an extra requirement is that the polarization directions of incident plane waves should be perpendicular to each other.  We show in Fig.~\ref{fig2}(g) (simulation) and Fig.~\ref{fig2}(h) (CDT) the scattering patterns on the $\textbf{x-z}$ and $\textbf{x-y}$ planes for the two scenarios in Figs.~\ref{fig2}(e)-(f), with $\phi_p=20^{\circ}$ and $110^{\circ}$, respectively (it could be arbitrary angles as long as the two polarizations are perpendicular). Both sets of results agree with each other and can confirm the invariance of the patterns under the duality transformation. The corresponding 3D scattering patterns (simulation) shown in Figs.~\ref{fig2}(e)-(f) render a further confirmation of such invariance.

\begin{figure}[tp]
\centerline{\includegraphics[width=8.8cm]{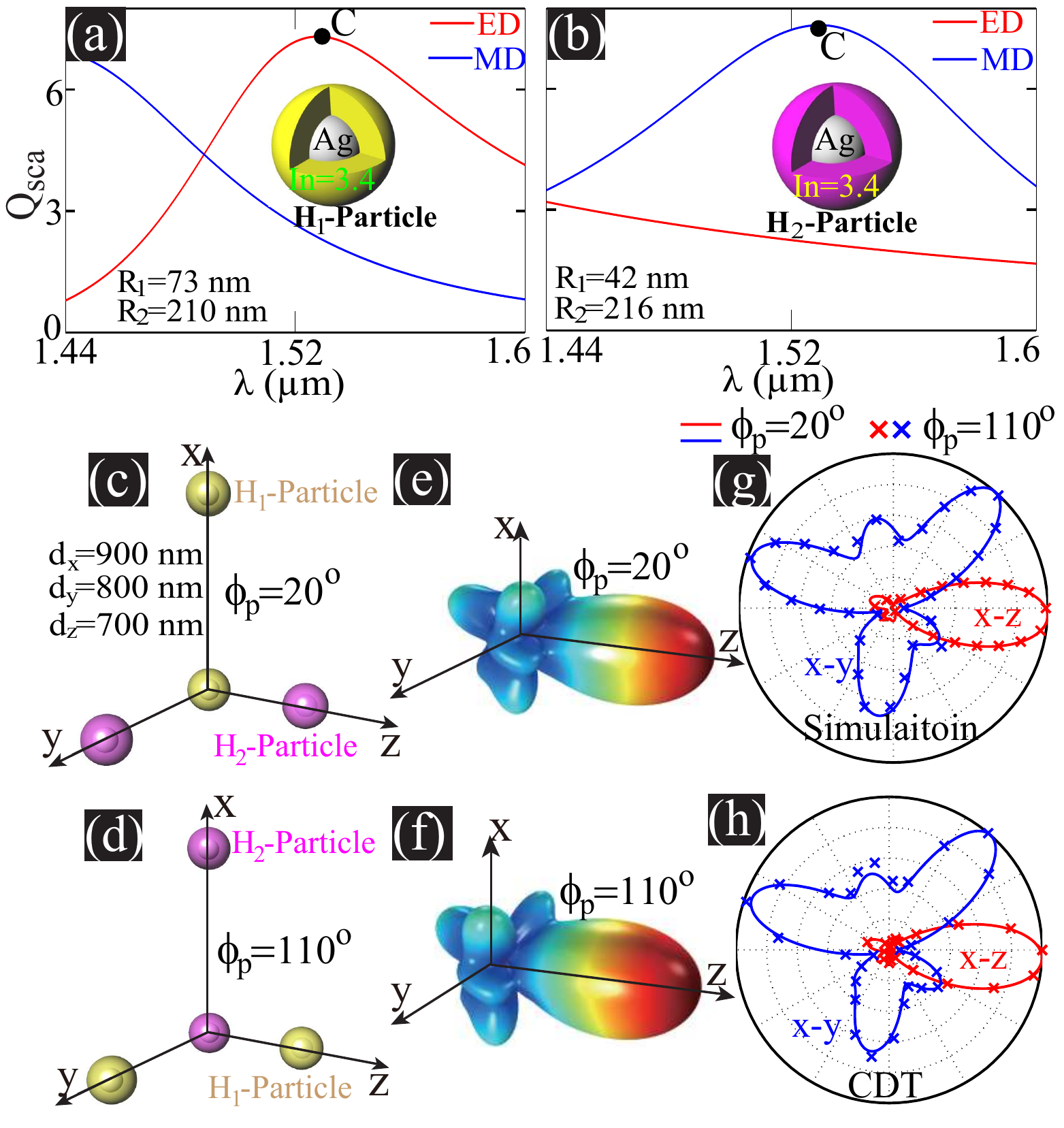}} \caption{(a) \& (b) Scattering efficiency spectra (dipolar contributions from ED and MD) for the Ag core-dielectric (In=$3.4$) shell spherical particles. At the common marked point $\lambda_\mathbf{C}=1528$~nm, both $\mathbf{H_1}$-\textbf{Particle} ($R_1=73$~nm and $R_2=210$~nm) and  $\mathbf{H_2}$-\textbf{Particle} ($R_1=42$~nm and $R_2=216$~nm) support not only electric but also magnetic moments, as shown respectively in (a) and (b), and they constitute a dual-pair. (c) \& (d) Two scattering configurations that are related to each other by a duality transformation of $\beta=\pi/2$ (interchanges between $\mathbf{H_1}$-\textbf{Particles} and $\mathbf{H_1}$-\textbf{Particles}), with $\phi_p=20^{\circ}$ and $110^{\circ}$ respectively. The two clusters shown are the same as that in Fig.~\ref{fig1}(c), except that the \textbf{D}-\textbf{Particles} are replaced by either $\mathbf{H_1}$-\textbf{Particles} or $\mathbf{H_2}$-\textbf{Particles}. The 2D scattering patterns (simulation and CDT results) on both $\textbf{x-z}$ and $\textbf{x-y}$ planes at $\lambda_\mathbf{C}$ are shown in (g) and (h), and the corresponding 3D scattering patterns are shown in (e) and (f).}\label{fig3}
\end{figure}

The scattering efficiency spectra for a more general dual-paired particles are summarized in Figs.~\ref{fig3}(a) and (b), which are identified respectively as $\mathbf{H_1}$-\textbf{Particle} ($R_1=73$~nm and $R_2=210$~nm) and $\mathbf{H_2}$-\textbf{Particle} ($R_1=42$~nm and $R_2=216$~nm). Here \textbf{H} means that the particles are hybrid, supporting both electric and magnetic moments ($a_{1}\neq 0,~ b_{1}\neq 0$). As is indicated  by Figs.~\ref{fig3}(a) and (b), the two constitute a dual-pair since $a_{1}(\mathbf{H_1}) =b_{1}(\mathbf{H_2})$ and $b_{1}(\mathbf{H_1}) =a_{1}(\mathbf{H_2})$  at the marked wavelength $\lambda_\mathbf{C}=1528$~nm. Similar to what is shown in Fig.~\ref{fig2}, the scattering configurations shown in Figs.~\ref{fig3}(c) and (d) can be converted to each other through a duality transformation, and thus the scattering patterns are the same for both scenarios, as have been verified by the corresponding 2D scattering patterns in Figs.~\ref{fig3}(g)-(h) and 3D scattering patterns (simulation) in Figs.~\ref{fig3}(e)-(f). In Figs.~\ref{fig2} and \ref{fig3}, despite that different dual-pairs of particles are employed, the same geometric (location) parameters (also the same as that in Fig.~\ref{fig1}) and polarization angles are chosen. Nevertheless, we have to emphasize that our conclusions of polarization pattern invariance is independent of the geometric distributions of the particles and the polarization directions of the incident waves, as long as they are perpendicular to each other for the pair connected by a duality transformation of $\beta=\pi/2$. Similar to those shown in Figs.~\ref{fig1}(d) and (e), the minor deviations and discrepancies in Figs.~\ref{fig2}(g) and (h), and in Figs.~\ref{fig3}(g) and (h) are induced either by the quadrupolar contributions and that the two particles replacing each other do not exactly constitute a dual-pair.

\section{Rotationally-symmetric scattering patterns with zero backward components induced by duality-rotation symmetry}
\label{section4}

As a first step, we refer to Fig.~\ref{fig4} to clarify the concept of duality-rotation symmetry. Here for simplicity, we have located all particles (represented by colored disks) on the $\mathbf{x-y}$ plane and two disks of different colors (yellow and green) constitute a dual-pair. Our arguments below can be applied to particles of more complicated 3D distributions and can certainly accommodate more different dual-pairs.  The incident wave (represented by $\mathbf{E}_0$ and $\mathbf{H}_0$) is propagating along $\mathbf{z}$. In Fig.~\ref{fig4}(a) we show two successive operations on an arbitrary particle cluster: a duality transformation T($\beta$) and a rotation operation of opposite rotation angle R($-\beta$) with respective to the incident wave vector \textbf{k} (for this specific demonstration $\beta=\pi/3$). As has been discussed in Section~\ref{section2}, for the incident wave, T($\beta$) is equivalent to a 2D geometric rotation along \textbf{k} and thus T($\beta$)=R($\beta$). As a result, the two successive operations will always bring the incident wave back to itself, as R(-$\beta$)R($\beta$)=\textbf{I} is basically the identity matrix. In contrast, for the scattering particles, the duality transformation is equivalent to an interchange between the dual-partners. As a result, the two successive operations will not necessarily bring the particle cluster back to itself, except for some special distributions.  The scattering configuration shown in Fig.~\ref{fig4}(b) exhibits the so called combined duality-rotation symmetry. To be more specific, what is shown here is only duality-(6-fold) rotation symmetry for the cluster: R$_6$($-\pi/3$)T($\pi/3$)=\textbf{I}.  As will be discussed below, more general duality-(n-fold) rotation symmetry with $n\geq 3$ can be found: R$_n$($-\beta$)T($\beta$)=\textbf{I}, where $\beta$ could be an integer number of $2\pi/n$.

\begin{figure}[tp]
\centerline{\includegraphics[width=8cm]{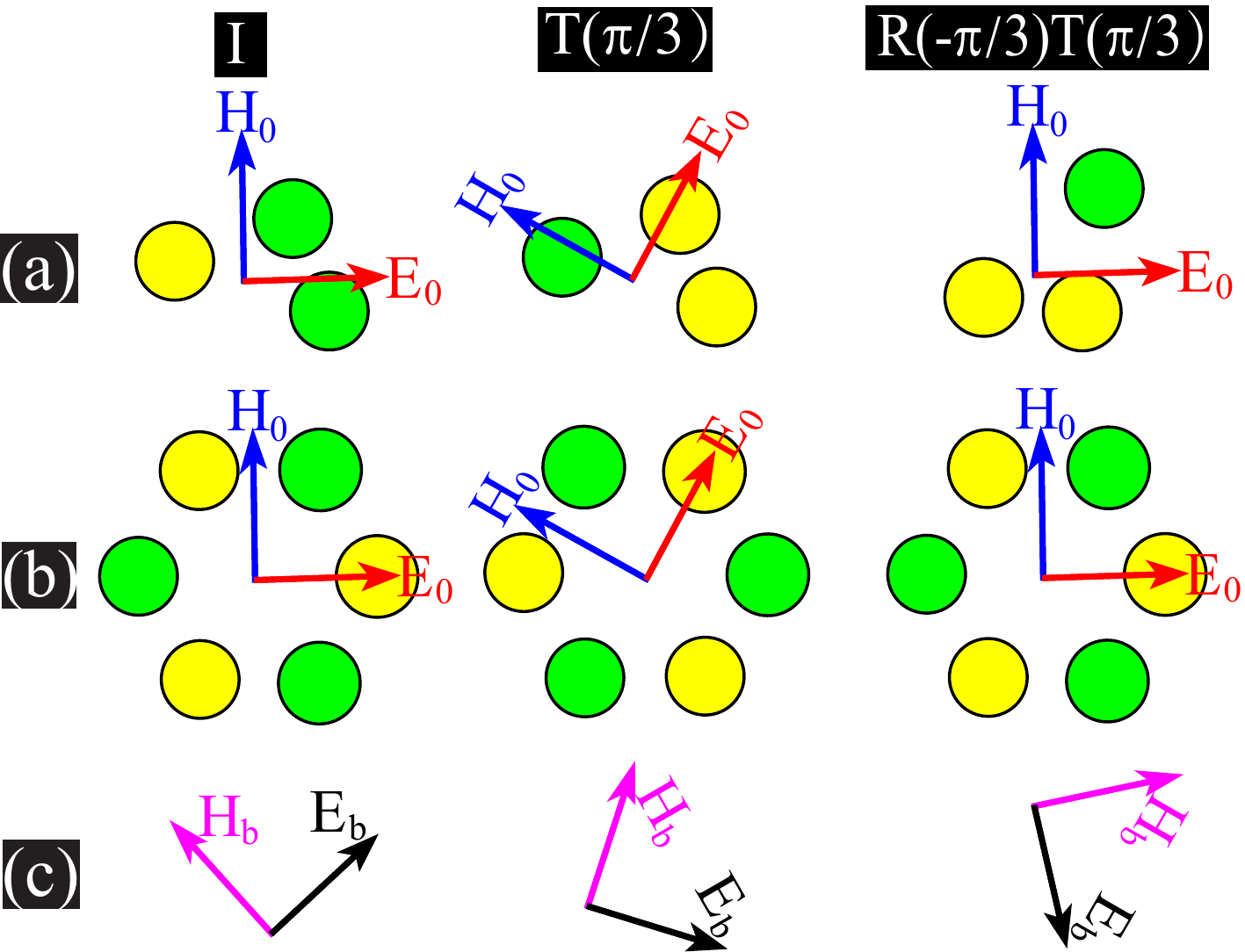}} \caption{The evolutions of incident waves ($\mathbf{E_0}$ and $\mathbf{H_0}$) and the scattering particles under a duality transformation ($\beta=\pi/3$) and a following rotation operation with opposite rotation angle ($\beta=-\pi/3$). The disks in yellow and green constitute a dual-pair. The cluster in (b) exhibit duality-(6-fold) rotation symmetry while that in (a) does not. The evaluations of the backward scattered waves ($\mathbf{E_b}$ and $\mathbf{H_b}$) are shown in (c).}
\label{fig4}
\end{figure}

As a next step, we discuss how the duality-rotation symmetry can constrain the overall scattering pattern, which is represented by \textbf{P}. Since a combined duality transformation and a geometric rotation of opposite angle would bring both the incident wave and the cluster of duality-rotation symmetry back to themselves [see Fig.~\ref{fig4}(b)], the scattering pattern would also be invariant under two such successive operations: R$_n$($-\beta$)T($\beta$)\textbf{P}=\textbf{P}. At the same time, as has been clarified in Section~\ref{section-3}, the scattering pattern is invariant under a duality transformation: T($\beta$)\textbf{P}=\textbf{P}. This directly leads to $R_n$($-\beta$)\textbf{P}=\textbf{P}, which by definition means that the scattering pattern exhibits n-fold rotation symmetry along \textbf{k}.

Then we proceed to study how the duality-rotation symmetry eliminates the backward scattering and thus leads to first Kerker scattering~\cite{Kerker1983_JOSA,LIU_2018_Opt.Express_Generalized}.  As has been discussed in Sections~\ref{section2} and \ref{section-3}, for incident waves and far-field scattered waves  in free space with perpendicular electric and magnetic fields, the duality transformation is identical to a geometric rotation with the rotation axis parallel to the Poynting vector. For the incident wave, it is shown above that T($\beta$)=R($\beta$); while for the backward scattered waves (represent by \textbf{F}$_b$), since the Poynting vector direction is now reversed compared to the incident wave, we have T$_b$($\beta$)=R(-$\beta$). Consequently, the backward wave would be converted by the successive duality transformation and rotation as:  T$_b$($\beta$)R(-$\beta$)\textbf{F}$_b$=R(-2$\beta$)\textbf{F}$_b$.  At the same time, the duality-rotation symmetry secures that T$_b$($\beta$)R(-$\beta$)\textbf{F}$_b$=\textbf{F}$_b$. The two points contradict each other unless \textbf{F}$_b$=0 (other possibilities like $\beta=\pi$ are physically trivial), which means the backward scattering is eliminated by sole duality-rotation symmetry. For a more intuitive understanding, in Fig.~\ref{fig4}(c) we show the evolutions of the backward scattered electric and magnetic fields (\textbf{E}$_b$ and \textbf{H}$_b$) under the successive operations, where without losing generality the initial fields are indicated by vectors of arbitrarily chosen orientations. It is shown that for the duality-rotation symmetric case [such as that shown in Fig.~\ref{fig4}(b)], the backward scattered waves would be brought back to itself after the two operations, which is not possible generally [see the final field orientations in Fig.~\ref{fig4}(c)] unless the backward scattering is zero.

\subsection{Scatterings by self-dual particle clusters with duality-rotation symmetry}

We have proved above the duality-rotation symmetry directly leads to rotationally symmetric scattering patterns of zero backward scattering, and now we aim to demonstrate such properties with the previously employed core-shell particles. We start with self-dual particle [\textbf{D}-\textbf{Particle} in Fig.~\ref{fig2}] clusters, of which sole n-fold rotation symmetry would directly lead to duality-(n-fold) rotation symmetry, since such clusters are still themselves under duality transformations of arbitrary angle $\beta$. As a result, duality-(n-fold) rotation symmetries of all $n\geq 3$ are supported. We show in Fig.~\ref{fig5}(a) three \textbf{D}-\textbf{Particle} clusters with n-fold ($n=3,~4,~5$) rotation symmetry: \textbf{D}-\textbf{Particles} located on the vertices of regular n-sided polygons of side length $d$ on the \textbf{x-y} plane. Here we have located all particles on the same plane, but our conclusions are valid for general 3D distributions as long as rotation symmetry is preserved. Since the scattering properties to be verified are independent on polarizations of incident waves, we randomly select $\phi_p=10^{\circ}$ for the following demonstrations.

\begin{figure}[tp]
\centerline{\includegraphics[width=8.5cm]{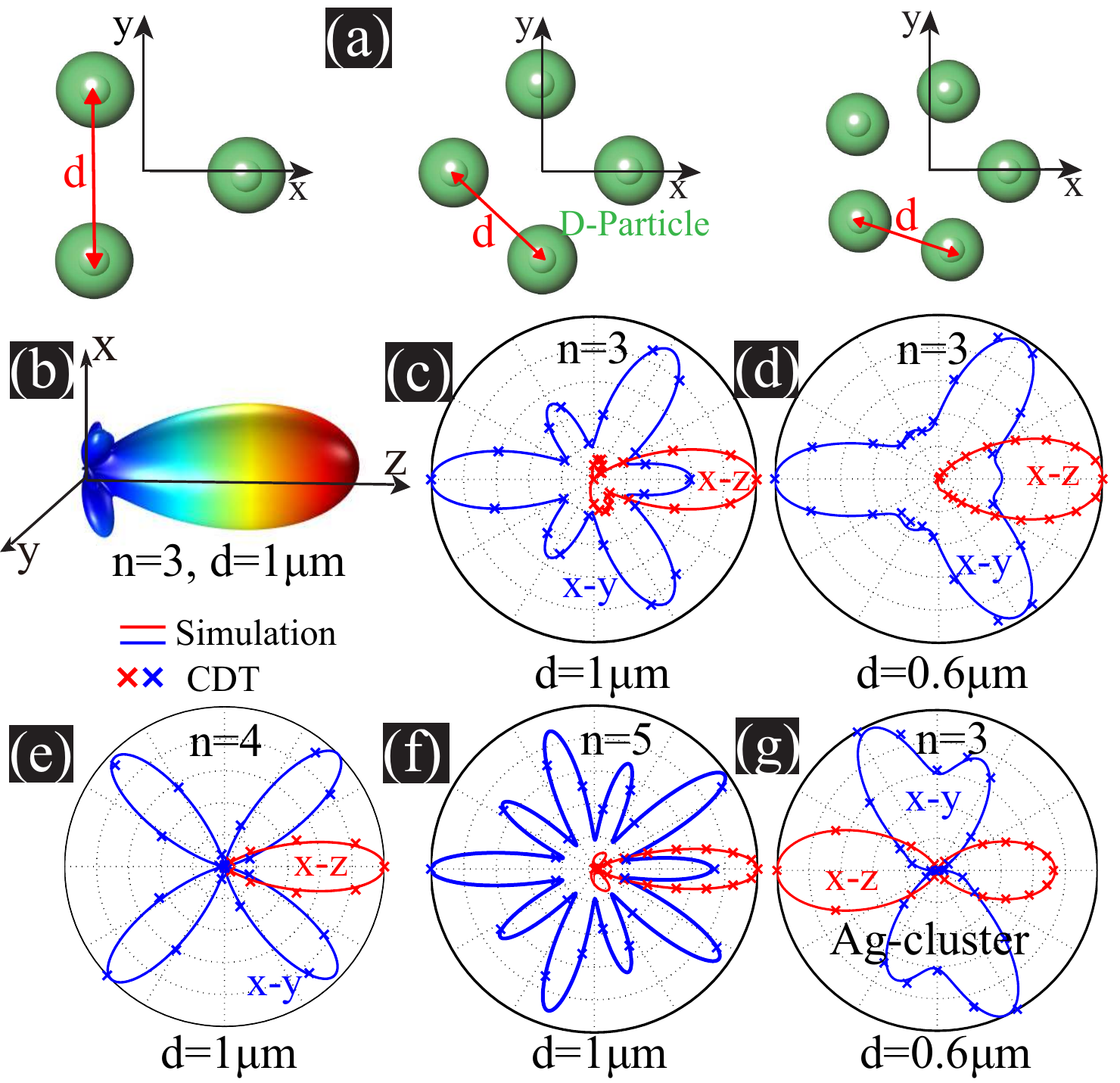}} \caption{\small (a) \textbf{D}-\textbf{Particle} clusters on the \textbf{x-y} plane, with all particles located on  vertices of regular n-sided polygons ($n=3,4,5$) of side length $d$.  For $n=3,~d=1000$~nm scenario, the 3D and 2D scattering patterns  are shown in (b) and (c), respectively. Other 2D patterns are shown: (d) for $n=3$, $d=600$~nm; (e) \& (f) for $d=1000$~nm and $n=4,~5$, respectively; and (g) for $n=3,~d=600$~nm, with the dielectric shell of the  \textbf{D}-\textbf{Particle} removed. For all scattering patterns, the corresponding polarization angle is $\phi_p=10^{\circ}$ and wavelength is $\lambda_\mathbf{A}=1128$~nm. The 2D patterns include scatterings on both $\textbf{x-z}$ and $\textbf{x-y}$ planes, with both simulation and CDT results shown.}
\label{fig5}
\end{figure}

We show the 3D scattering pattern from simulation ($n=3$, $d=1000$~nm) in Fig.~\ref{fig5}(b), which as expected exhibit 3-fold rotation symmetry along $\textbf{z}$  and zero backward scattering. Both features are further confirmed by the 2D scattering patterns on the \textbf{x-y} and \textbf{x-z} planes in Fig.~\ref{fig5}(c), where both simulation and CDT results are shown. As has already been argued, those properties are symmetry protected and thus immune to different coupling strengths between the particles, which can be verified by the patterns of another cluster of a different inter-particle distance ($n=3$, $d=600$~nm), as shown in Fig.~\ref{fig5}(d). As a result, the first Kerker scattering shown here is fundamentally different from the other ones previously demonstrated with particle clusters, where the zero backscattering is highly sensitive to couplings and thus the Kerker condition can be easily violated (see Review~\cite{LIU_2018_Opt.Express_Generalized} and references therein).  Such robust responses are also manifest in periodic structures with the required symmetry, as will be demonstrated in Fig.~\ref{fig7}. Rotationally symmetric scattering patterns with zero backward scattering can be also observed for other particle clusters with rotation symmetries, as shown in Fig.~\ref{fig5}(e) ($n=4$, $d=1000$~nm) and Fig.~\ref{fig5}(f) ($n=5$, $d=1000$~nm). Our arguments can be directly extended to the scenario of continuous rotation symmetry ($n=\infty$), which would directly result in azimuthally symmetric ($\phi$-independent) scattering patterns with zero backward scattering, as has already been demonstrated with an individual self-dual core-shell spherical particle~\cite{Liu2012_ACSNANO}.

It might be taken for granted that the n-fold rotational symmetry of scattering patterns is solely induced by the rotational symmetry of the cluster, having nothing to do with the overall duality-rotation symmetry. This is incorrect, since when the incident wave is taken into consideration, the rotation symmetry of the whole scattering configuration is broken. In Fig.~\ref{fig5}(g) we show the 2D scattering patterns of the cluster ($n=3$, $d=1000$~nm) consisting of four Ag spheres [\textbf{D}-\textbf{Particles} with the dielectric shells removed]. Now the duality-rotation symmetry of the cluster is broken with the rotation symmetry maintained. As shown by Fig.~\ref{fig5}(g), however, neither a rotationally symmetric pattern nor zero backward scattering is observed. We note that the rotation symmetry of the scattering patterns are not ideal in Figs.~\ref{fig5}(c)-(f), simply because the [\textbf{D}-\textbf{Particles} are not exactly self-dual. Moreover, there are also discrepancies between results obtained through simulation and CDT, as the latter has fully neglected the contributions from higher-order multipoles. Similar minor deviations can be also observed in the following studies shown in Figs.~\ref{fig6} and \ref{fig7}.

It is worth noting the scattering properties  we have covered here have been partially (symmetry-protected zero backward scattering only) studied in Ref.~\cite{FERNANDEZ-CORBATON_2013_Opt.ExpressOE_Forwarda}, where the discussions are focusing on bulk magnetic materials ($\mu \neq 1$). In contrast, our demonstrations here are fully based on realistic nonmagnetic particles that support optically-induced magnetic responses. Moreover, as far as we understand, our proof of such a symmetry-protected Kerker scattering is quite different and much simpler, without replying on the principle of helicity conservation that plays an essential role in Ref.~\cite{FERNANDEZ-CORBATON_2013_Opt.ExpressOE_Forwarda}.

\subsection{Scatterings by non-self-dual particle clusters with duality-rotation symmetry}

\begin{figure}[tp]
\centerline{\includegraphics[width=8.5cm]{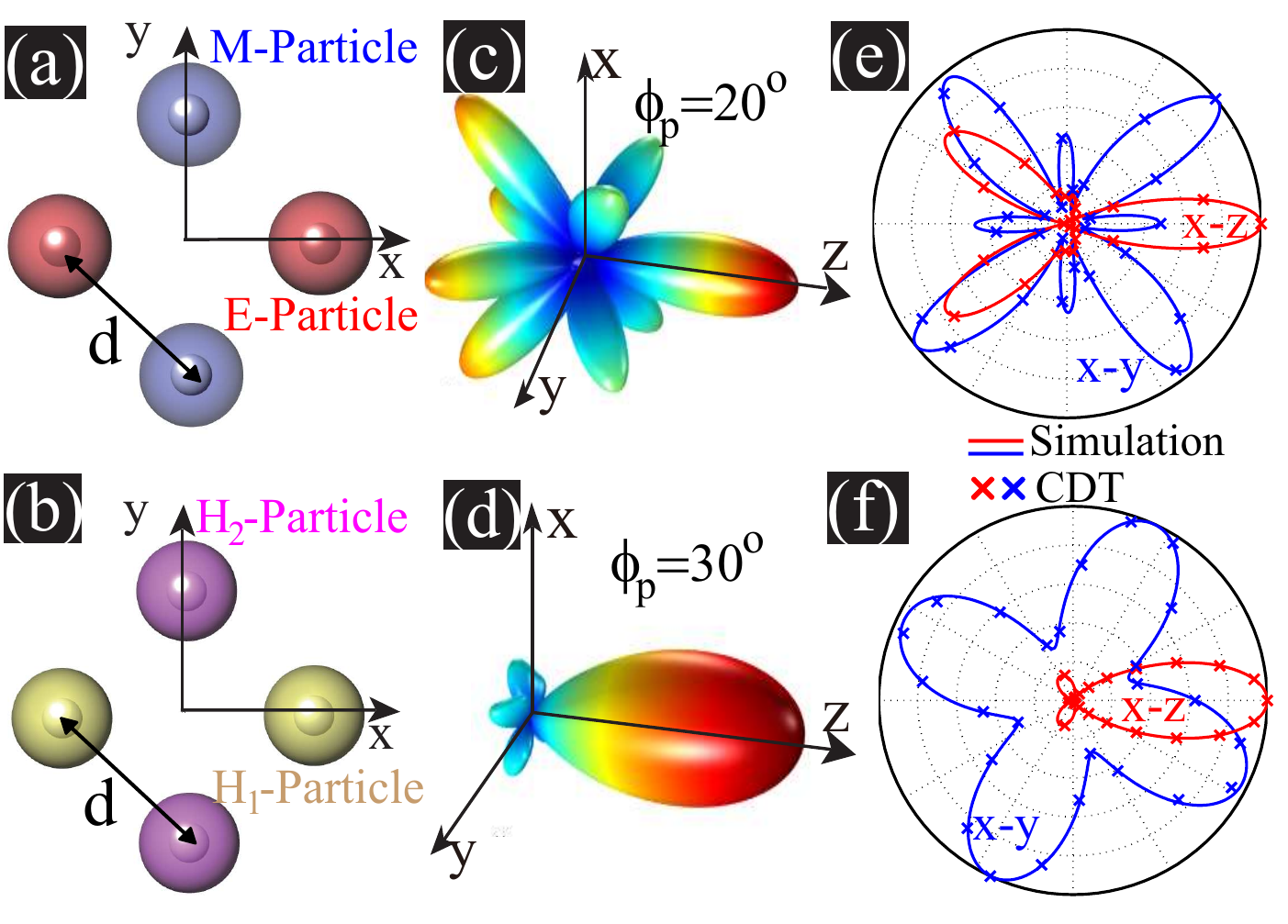}} \caption{\small Square (side length $d$) clusters on the \textbf{x-y} plane with identical particles locating at diagonal positions, consisting of \textbf{E, M-Particles} in (a), and $\mathbf{H_1,H_2}$-\textbf{Particles} in (b). The 3D scattering patterns are shown respectively in (c) and (d), and 2D patterns (on $\textbf{x-z}$ and $\textbf{x-y}$ planes, with both simulation and CDT results) respectively in (e) and (f). The corresponding polarization angle and wavelength are $\phi_p=20^{\circ}$, $\lambda_\mathbf{B}=920$~nm in (c) and (e); $\phi_p=30^{\circ}$, $\lambda_\mathbf{C}=1528$~nm in (d) and (f).}
\label{fig6}
\end{figure}

As has been explained in Section~\ref{section2}, for non-self-dual particle clusters with isotropic responses of no magneto-electric couplings, only the duality transformation of $\beta=\pi/2$ can be implemented. The symmetry condition R$_n$($-\beta$)T($\beta$)=\textbf{I} can be satisfied only for $n=4$, indicating that only duality-(4-fold) rotation symmetry is allowed for such non-self-dual particle clusters.  We show in Figs.~\ref{fig6}(a) and (b) two such square clusters ($d=1000$~nm), consisting of the dual-pairs of \textbf{E, M-Particles} (refer to Fig.~\ref{fig2}) and $\mathbf{H_1,H_2}$-\textbf{Particles} (refer to Fig.~\ref{fig3}), respectively. For both square clusters: the 3D scattering patterns (simulation) are shown respectively in Figs.~\ref{fig6}(c) and (d); the 2D scattering patterns on the \textbf{x-z} and \textbf{x-y} planes are summarized respectively in  Figs.~\ref{fig6}(e) and (f) (simulation and CDT). As is shown, both properties of 4-fold rotationally symmetric scattering patterns and zero backward scatterings can be observed. We further note that similar results (related to the zero backward scattering property only) have also been partially presented in a recent work of Ref.~\cite{MOHAMMADIESTAKHRI_2020_Phys.Rev.Lett._Electromagnetic}. This previous study is based on ideal bulk magnetic materials ($\mu \neq 1$) or perfect magnetic conductors, which is less practical, especially in the optical regime.  Moreover, we have placed our study in a much broader background of duality-rotation symmetry, showing that what is presented in Ref.~\cite{MOHAMMADIESTAKHRI_2020_Phys.Rev.Lett._Electromagnetic} is a special case of such general symmetry that can intrinsically eliminate the backward scattering.

\subsection{Duality-rotation symmetry induced reflection suppression for periodic structures}

\begin{figure}[tp]
\centerline{\includegraphics[width=8.2cm]{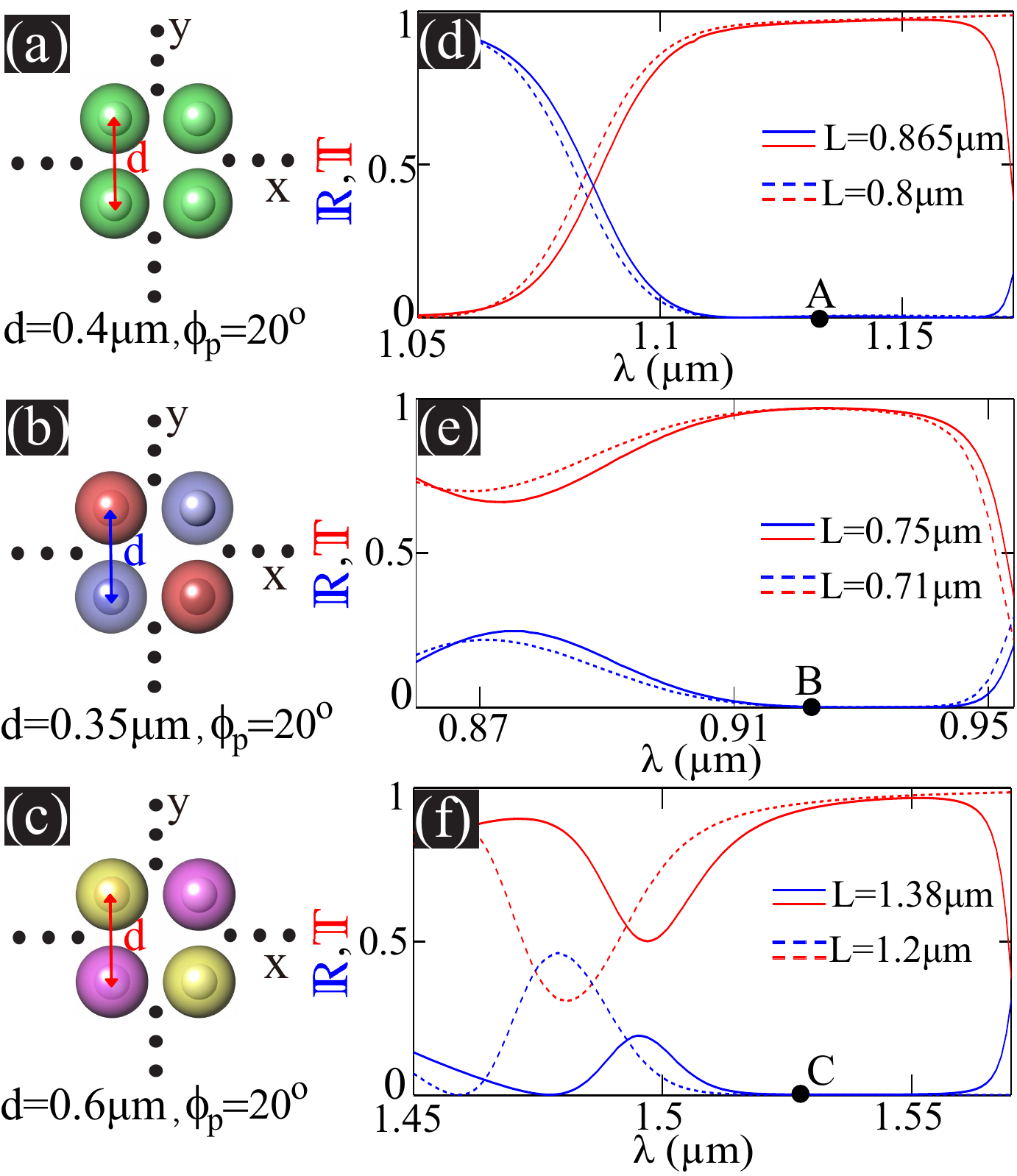}} \caption{\small Periodic square lattices of perodicity \textbf{L} with the clusters of duality-(4-fold) rotation symmetry as the unit-cells: (a) \textbf{D-Particle} cluster with $d=400$~nm and \textbf{L}$=800$~nm, $865$~nm; (b) \textbf{E, M-Particle} cluster with $d=350$~nm and \textbf{L}=$710$~nm, $750$~nm; and (c) $\mathbf{H_1,H_2}$-\textbf{Particle} cluster with $d=600$~nm and \textbf{L}=$1200$~nm, $1380$~nm. For the periodic lattices, two sets of reflection and transmission spectra with $\phi_p=20^{\circ}$ are shown in: (d) \textbf{L}=$800$~nm, $1000$~nm; (e) \textbf{L}$=710$~nm, $750$~nm; and (f) \textbf{L}$=1200$~nm, $1380$~nm. The reflection is zero at not only the marked points: $\lambda_\mathbf{A}=1128$~nm, $\lambda_\mathbf{B}=920$~nm and  $\lambda_\mathbf{C}=1528$~nm, but also across a broad spectral regime beyond them.}
\label{fig7}
\end{figure}

The discussions above about finite clusters can be naturally extended to infinite periodic structures as far as the required symmetry is maintained~\cite{LIU_2017_ACSPhotonics_Beam,BABICHEVA_Laser&PhotonicsReviews_resonant}.  Here based on the three square clusters (inter-particle distance $d=400$~nm, $350$~nm, $600$~nm, respectively) studied in Figs.~\ref{fig5} and \ref{fig6} as unit-cells, we construct periodic square lattices (periodicity \textbf{L}) shown in Figs.~\ref{fig7} (a)-(c).  The reflection ({\color{blue}$\mathbb{R}$}) and transmission ({\color{red}$\mathbb{T}$}) spectra of all three lattices are shown in Figs.~\ref{fig7} (d-f)(obtained through COMSOL simulations), where for each scenario two sets of spectra for different \textbf{L} are included. As has been discussed in Sec.~\ref{section4}, the reflection suppression feature is polarization independent and thus we arbitrarily choose incident angle $\phi_p=20^{\circ}$ for our demonstrations. It is shown that zero reflection is observed respectively at $\lambda_\mathbf{A}=1128$~nm, $\lambda_\mathbf{B}=920$~nm and  $\lambda_\mathbf{C}=1528$~nm, for both sets of spectra. Moreover, this feature is preserved across a broad spectral range rather than only at the single designed frequency. This is due to the fact that the paired particles remain more or less to be dual-partners [According to Figs.~\ref{fig1}-\ref{fig3}, $(a_{1},b_{1}) \rightarrow (b_{1},a_{1})$ can be matched approximately between the pairs] and thus preserve the duality-rotation symmetry at broad spectral regimes beyond the designed point. This feature can potentially lead to broadband applications such as robustly enhanced transmissions, especially when low-dispersive materials rather than metals are employed. Here we have confined our demonstrations to square lattices. Nevertheless, as has been revealed previously, for the self-dual particle clusters, duality-rotation symmetries of order $3$ and $6$ are also allowed. It means that self-dual periodic lattices of $3$- and $6$-fold rotational symmetry can be achieved, with duality-rotation symmetry protected eliminated reflections.

\section{Conclusions and Discussions}

In conclusion, we have merged the fundamental concepts of optically-induced magnetism and electromagnetic duality, which brings extra flexibilities for the scattering manipulations of nonmagnetic particle systems, and produce robust symmetry-protected scattering properties. It is discovered that under duality transformations, the far-field scattering patterns are invariant. For self-dual particle clusters, this secures that the scattering patterns are independent on the incident polarizations with fixed incident directions. Based on this discovery, we further reveal that any scattering systems of combined duality-(n-fold) rotation symmetry would also exhibit n-fold rotationally symmetric scattering patterns, with the backward scatterings automatically eliminated (first Kerker scattering). We have verified all those properties with nonmagnetic core-shell particles, liberating the conventional electromagnetic duality related fundamental researches and applications from the rather rare intrinsically magnetic materials.

In this work, concerning specific demonstrations related to the duality-rotation symmetry (see Section~\ref{section4}), for simplicity we have located all particles on the same plane perpendicular to the incident direction, and the clusters also exhibit extra mirror symmetry. It is worth mentioning that the revealed scattering features are solely induced by duality-rotation symmetry, which would be fully preserved even when the mirror symmetry is broken. Also for the non-self-dual particle clusters, we have confined our study to isotropic responses without magneto-electric coupling, which permits only the duality transformation of $\beta=\pi/2$ and thus only duality-(4-fold) rotation symmetry is allowed. It is expected that when bi-isotropic or even bi-anisotropic materials are involved~\cite{TRETYAKOV_2003__Analytical,FERNANDEZ-CORBATON_2013_Phys.Rev.B_Role}, non-self-dual particle clusters can exhibit other duality-(n-fold) rotation symmetries [see \textit{e.g.} Fig.~\ref{fig4} (b) where $n=6$], which enables constructions of more general periodic or quasi-periodic structures [beyond what are shown Figs.~\ref{fig7} (b) and (c)], such as hexagonal lattices with 6-fold rotational symmetry.

Here we have confined our study to dipolar particles, which by no means implies that our study is of very limited value, since in principle all particles can be reduced by numerical discretizations into dipolar moments~\cite{Mulholland1994_Langmuir,Merchiers2007_PRA,DRAINE_JOSAA_discretedipole_1994}. When the discretizations are not implemented, each particle can support higher-order multipolar moments at sufficiently higher frequencies, for which the dual-partner might not be uniquely defined from interchanging Mie coefficients $a_n$ and $b_n$. This is an interesting open question, as far as we know, since magnetism emerges from not only magnetic multipoles, but also from electric multipoles of higher orders~\cite{CHO_2008_Phys.Rev.B_Contribution,LIU_Phys.Rev.Lett._generalized_2017}. The demonstrations in this work are all based on Ag core-dielectric shell particles, which can easily be extended to homogeneous all-dielectric particles that also support both electric and magnetic moments. The scattering properties we have obtained are symmetry protected, independent of specific local geometric and optical parameters, and immune to perturbations which preserves the required symmetry. Considering the ubiquitous roles of particle scatterings all across different branches of photonics, we believe our results can bring new opportunities for not only fundamental researches but also practical applications in various optical devices that desire optically stable functionalities.

\section*{acknowledgement}
We acknowledge the financial support from National Natural Science
Foundation of China (Grant No. 11874026, 11404403 and 11874426), and the Outstanding Young Researcher Scheme of National University of Defense Technology.

Q. Yang and W. Chen contributed equally to this work.

\end{document}